\documentclass[11pt,twoside]{article} 
\usepackage{asp2004}
\usepackage{graphicx}
\usepackage{lscape} 

\begin{document} 

\title{White Dwarf Light Curves: Constraining Convection and Mode
  Identification Using Non-sinusoidal Light Curves} \author{M. H.
  Montgomery} \affil{Dept. of Astronomy, University of Texas, Austin,
  TX, 78712, USA}

\begin{abstract} 
  
  In this paper we show how the formalism of Wu \& Goldreich developed
  for weakly nonlinear white dwarf pulsations can be extended to the
  fully nonlinear case. Due to the computational simplification this
  affords over the simulations of \citet{Brickhill92a} and
  \citet{Ising01}, we are able to obtain the first ever \emph{fits} of
  observed light curves. These fits allow us to obtain not only mode
  identifications but also to constrain the average depth of the
  convection zone \emph{and} its sensitivity to changes in $T_{\rm
    eff}$. We are thus able to probe the fundamental physics of
  convection in these objects.

\end{abstract}

\section{Astrophysical Context}

In the past decade, the seismology of pulsating stars in general
(``asteroseismology''), and of white dwarfs in particular, has
produced remarkable results concerning the parameters and internal
structure of stars. Besides determining very precise luminosities and
therefore distances to many of these stars
\citep[``asteroseismological parallax'',
e.g.][]{Bradley94a,Winget94,Winget91}, we have placed constraints on
many aspects of their internal structure, such as their helium and
hydrogen surface layer masses \citep[e.g.][]{Metcalfe00} and their
core carbon/oxygen profiles \citep{Metcalfe03}; in addition, we have
provided empirical evidence that crystallization in cooling white
dwarfs does in fact occur \citep{Metcalfe04,Montgomery99a,Salpeter61}.
Finally, we have recently made fundamental progress in understanding
how the observed eigenfrequencies sample the radial structure of our
models \citep{Montgomery03}, and how the resulting ``core/envelope
symmetry'' means that our models are actually very sensitive to the
structure of the deep core.

However, all of this progress in the field of white dwarf
asteroseismology has involved the observation and modelling of their
linear, sinusoidal pulsations. While this program of investigation has
been very fruitful, it is somewhat ironic that about 2/3 of all white
dwarf pulsators are large-amplitude and nonlinear.  This sets them
apart from most other classes of non-radially pulsating stars, which
are usually small amplitude (e.g.\ $\delta$~Scuti, $\beta$~Cephei,
$\gamma$~Doradus, and solar-like pulsating stars), and places them
with the large-amplitude radial pulsators, such as the RR Lyrae and
Cepheid stars.

These observed nonlinearities contain information about the physical
processes which produced them.  Along with previous researchers, we
believe these nonlinearities to be the result of the interaction of
the pulsations with the convection zone
\citep[e.g.][]{Brickhill92a,Brickhill92b,Wu01,Ising01}. Physically,
the convection zone varies in depth and therefore heat capacity during
a given pulsation cycle, with the result that it modulates the
(assumed) sinusoidal flux flowing into it from below, producing a
nonlinear flux variation at the photosphere. The amount by which the
depth of the convection zone varies is intimately tied to the physics
of convection in these stars, and therefore is a probe of this
physics.  Given that convection is even today somewhat unconstrained
in the modelling of these and other stars (the ``free'' parameter
$\alpha$), any progress we can make in providing independent
observational constraints is of definite value.

In the sections which follow, we show how a particular physical
approximation can be used to greatly simplify the description of the
pulsation/convection interaction in these stars. Using models based on
this, we apply this method to the observed pulsations of two different
stars. These fits to observed light curves, and the subsequent
derivation of convective and pulsation mode parameters, represent the
first such results for white dwarf pulsators, and provide important
tests and points of contact for both older
\citep[e.g.][]{Bohm-Vitense58} and more modern
\citep{Montgomery04a,Kupka99} theories of convection.

\section{The Model}

\subsection{Assumptions}

While we are simulating the same basic physical system described in
\citet{Brickhill92a}, the set of assumptions which we adopt allows us
to model and therefore fit light curves much more easily \citep[see
][]{Wu01}.  The fundamental assumptions of this model are that:
\begin{enumerate}
\item The flux perturbations beneath the convection zone are
  completely sinusoidal.
\item The convection zone is so thin that we may locally ignore the
  angular variation of the nonradial pulsations, i.e., we treat the
  pulsations locally as if they were radial.
\item The convective turnover timescale is so short compared to the
  pulsation periods that the convection zone can be taken to respond
  ``instantaneously''.
\item Due to the extreme sensitivity of the convection zone, only flux
  and temperature variations need to be considered, i.e., the
  large-scale fluid motions associated with the pulsations can be
  ignored.
\end{enumerate}
The third assumption is given weight by appeal to standard convection
models of these stars, which indicate that the time taken for a
convective fluid element to circulate from top to bottom of the
convection zone should be less than a second. Since the pulsation
periods of these stars are much longer than this, i.e., hundreds of
seconds, this appears to be a safe assumption.  While plausibility
arguments for the other assumptions may similarly be made, it is
perhaps better to see how well they allow us to model observed
light curves before deciding on their viability (``the proof is in the
pudding'').

\subsection{Energetics}

We assume energy balance in the sense that the energy radiated away at
the photosphere (the top of the convection zone) is equal to the
energy absorbed at its base \emph{minus} the energy absorbed by the
convection zone itself. Mathematically, we write
\[
\underbrace{F_{\rm phot}}_{\parbox{2.5em}{\begin{center} \footnotesize 
    \vspace*{-1em} energy output \end{center}}} = 
\underbrace{F_{\rm base}}_{\parbox{2.5em}{\begin{center} \footnotesize
    \vspace*{-1em} energy input \end{center}}} - 
\underbrace{\frac{d \tilde{Q}}{d
    t},}_{\parbox{3.5em}{\begin{center} \vspace*{-1em}
      \footnotesize energy absorbed
  \end{center}}}
\]
i.e., the flux emitted at the photosphere equals the flux incident at
the base of the convection zone minus the rate at which flux (energy)
is absorbed by the convection zone. Physically, this can be thought of
as the convection zone having a specific heat. For instance, in order
for its temperature to be raised by 200~K, say, it must absorb a given
amount of energy.

Due to the assumption that the convection zone is instantaneously in
quasi-static equilibrium, we can compute the term $d\tilde{Q}/dt$
purely in terms of static models. As it turns out, this energy
absorption rate depends only on the photospheric flux (i.e., $T_{\rm
  eff}$), leading to the following equation:
\begin{equation} 
F_{\rm phot} = F_{\rm base} + \tau_C \frac{d F_{\rm phot}}{dt},
\label{engy}
\end{equation}
where the new timescale $\tau_C \equiv \tau_C \mbox{\scriptsize
  $(F_{\rm phot})$}$ describes the changing heat capacity of the
convection zone as a function of the local photospheric flux. Thus, we
have reduced the problem to a first-order equation in time. However,
we must remember that although $F_{\rm base}$ is assumed to be
sinusoidal, it still does have a spatial dependence ($F_{\rm base}
\propto {\rm Re}[e^{i \omega t} Y_{\ell m}(\theta,\phi)]$), so
equation~\ref{engy} must be solved on a grid of points across the
visible surface of the star and the resulting fluxes added together to
calculate the observed light curve.  Fortunately, as a consequence of
the simplifications we have introduced, this problem is still
tractable.

In Figure 1, we show a theoretical calculation of how this timescale
$\tau_C$ is expected to vary as a function of $T_{\rm eff}$. For a
given value of $\alpha$, we see that $\tau_C$ changes by approximately
a factor of 1000 as the temperature goes from 12000~K (the observed
onset of WD pulsations) to 11000~K (the observed disappearance of
pulsations). Since a typical pulsator may have excursions in
temperature of several hundreds of degrees, we expect $\tau_C$ to vary
greatly during the pulsations, leading to nonlinear light curves.

\begin{figure}[bt]
  \centerline{\includegraphics[height=0.84\columnwidth,angle=-90]{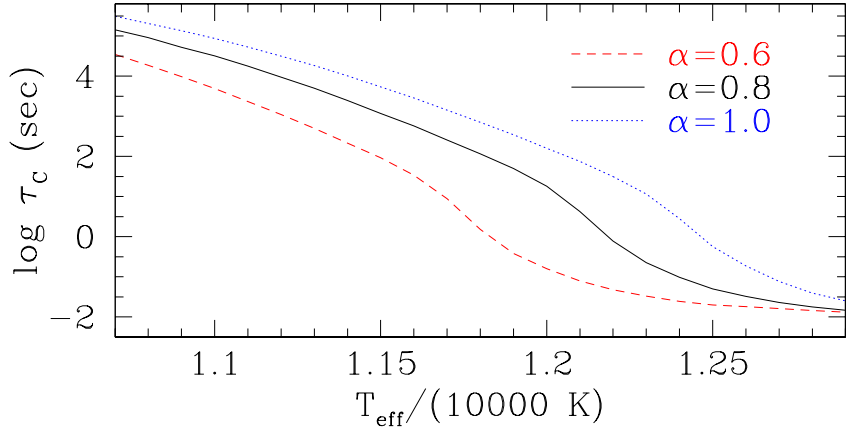}}
  \caption{Theoretical calculations of the thermal response 
    timescale of the convection zone, $\tau_C$, as a function of
    $T_{\rm eff}$. The local mixing-length theory of \citet{Bohm71}
    has been used, for the indicated values of $\alpha$.}
  \label{tauc}
\vspace*{-1em}
\end{figure}

\section{Light Curve Fitting}

Using Figure~\ref{tauc} as a guide, we see that although $\tau_C$ is a
strong function of temperature, its logarithm varies approximately
linearly with temperature. Since $\delta F/F = 4 \, \delta T/T$, $\log
\tau_C$ is also an approximately linear function of $F$, and since we
wish to make as few assumptions about convection as possible, we
choose
\[
\tau_C = \tau_0\, e^{a (F_{\rm phot}/F_{0}-1)},
\]
where $\tau_0$ and $a$ are unknown coefficients and $F_0$ is the
equilibrium flux.

In addition to the convective parameters $\tau_0$ and $a$, we fit the
amplitude and phase of the sinusoidal signal at the base of the
convection zone (the period is assumed already known from the
observations) as well as the inclination angle between the pulsation
axis and our line-of-sight.

\vspace*{-0.2em}
\subsection{The DBV PG1351+049}

\begin{figure}[t]
\centerline{\includegraphics[height=0.84\columnwidth,angle=-90]{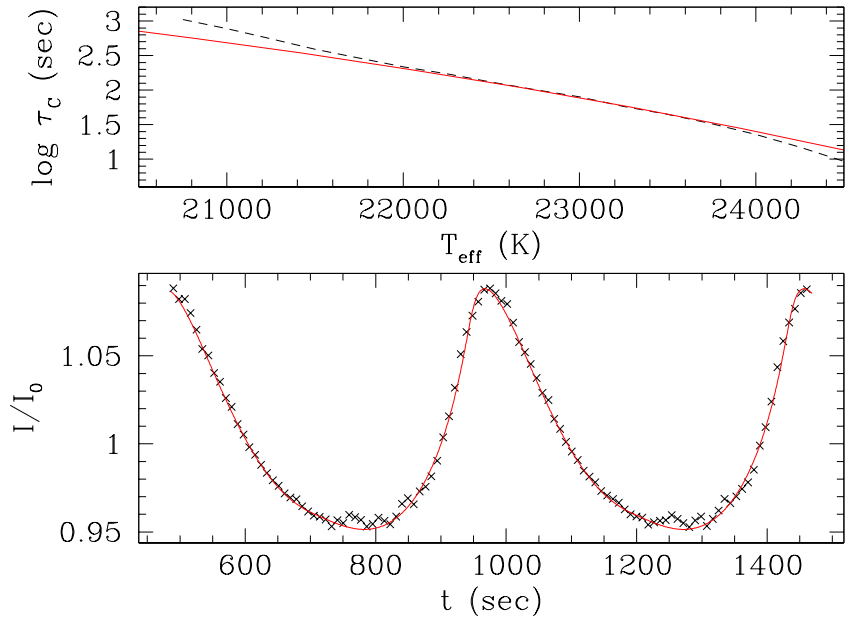}}
\vspace*{-0.3em}
  \caption{A fit to the observed pulse shape of the DBV star PG1351+049.}
  \label{pg1351}
\end{figure}

In 1995 data were acquired on the DBV star PG1351+049 as part of a
Whole Earth Telescope (WET) campaign
\citep[e.g.][]{Winget91,Winget94}. The resulting light curve, which
has been folded at the period of the mode, 489.335 s, is shown in the
lower panel of Figure~\ref{pg1351}, where the crosses represent the
data and the solid curve is our fit. We note that the fit is very good
and reproduces the features in the light curve quite well.  In the
upper panel we show the convective timescale which we derive from
these data (solid curve) as well as one computed using mixing-length
theory (dashed curve).  The fact that the slopes are similar is
encouraging, since the theoretical curve can be moved up or down
simply by tuning the value of $\alpha$, while its slope is less
sensitive to the choice of $\alpha$.

This star was observed again in May 2004 with the 2.1m telescope at
McDonald Observatory.  This allows us to make another fit, which
provides an important check since we would not expect the parameters
characterizing the star itself to have changed, i.e., the average
depth of the convection zone, $\tau_0$, and the inclination angle,
$\theta_{\rm i}$, should be the same, while the amplitude and phase of
the mode might very well be different. We summarize these two fits in
Table~1: we see that the amplitude of the mode has changed somewhat,
but $\theta_{\rm i}$ and $\tau_0$ remain the same, i.e., we appear to
be looking at the \emph{same star}.
\begin{table}[h!]
  \begin{center}
  Table1. \hspace{0em} Derived parameters for PG1351+049\\[0.4em]
\begin{tabular}{crcccc}
\hline\\[-0.8em]
year  & $\tau_0$ (sec) & $\theta_{\rm i}$ (deg) & Amp & $\ell$ & $m$ \\[0.2em]
\hline \\[-0.8em]
1995 &  99.95 & 59.4 & 0.325 & 1 & 0\\
2004 & 100.83 & 59.4 & 0.268 & 1 & 0 \\[0.1em]
\hline
\end{tabular}
  \end{center}
\end{table}

\subsection{The DAV G29-38}

\begin{figure}
 \centerline{\includegraphics[height=0.85\columnwidth,angle=-90]{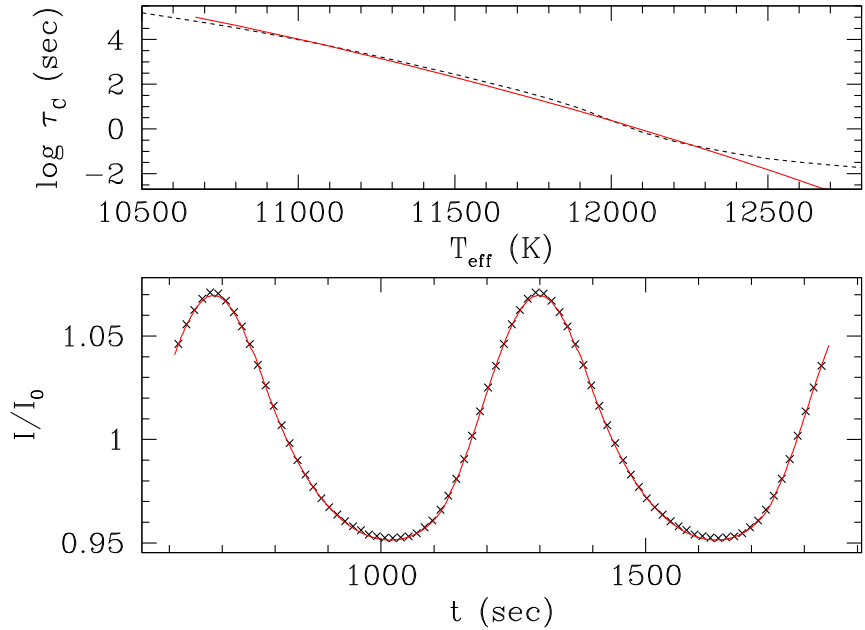}}
  \caption{A fit to the observed pulse shape of the DAV star G29-38.}
  \label{g29}
\end{figure}

The observations of the DAV G29-38, taken by S. Kleinman in 1988, are
displayed as the crosses in the lower panel of Figure~\ref{g29} (note:
the errors on this extremely high signal-to-noise data are actually
much smaller than the crosses). Our fit to this data is given by the
solid curve, which is nearly indistinguishable from the data except at
the maxima and minima, where slight differences can be seen.  In the
upper panel, we show the derived dependence of $\tau_C$ as a function
of $T_{\rm eff}$ (solid curve) as well as a theoretical curve computed
using mixing-length theory (dashed curve). We see that the slopes of
these curves are nearly identical over this region, again implying
that we really are measuring something fundamental about this star,
and about the physics of convection.

\section{Conclusions}

First, the fits we have obtained for the stars PG1351+049 and G29-38
are impressively good, implying that the physical assumptions which
have gone into our models are realistic. Second, our results seem to
make sense in terms of simple (and naive) theories of convection, and
they provide a means of calibrating such models. Third, the fits
provide an independent method for determining the $\ell$ and $m$
values of pulsation modes, something which is necessary in order to
use asteroseismology to learn about the deep stellar interior.

The ultimate goal of this research is to provide a detailed empirical
map of how convection works in these stars, which in terms of their
pulsations means mapping how $\tau_C$ varies as a function of $T_{\rm
  eff}$ and $\log g$. Since each pulsator essentially gives us the
value of $\tau_C$ \emph{and} its slope at a given $T_{\rm eff}$, by
applying this method to stars with temperatures spanning the
instability strips of the DAVs and DBVs we can derive this dependence
empirically.  In the future, this determination will provide an
important contact for more sophisticated and physical theories of
convection which go beyond mixing-length theory \citep[those without
``free parameters'', e.g.][]{Montgomery04a,Kupka02} and whose
applicability will be to stars throughout the HR diagram.

\acknowledgements{The author wishes to thank D. O. Gough, D. E.
  Winget, and D. Koester for useful discussions.}


\begin{thebibliography}{}

\bibitem[{B{\" o}hm-Vitense}(1958)]{Bohm-Vitense58}
{B{\" o}hm-Vitense}, E., 1958, Zeitschrift f\"ur Astrophysik, 46, 108

\bibitem[{B\"ohm} \& {Cassinelli}(1971)]{Bohm71}
{B\"ohm}, K.~H., \& {Cassinelli}, J., 1971, \aap, 12, 21

\bibitem[{Bradley} \& {Winget}(1994)]{Bradley94a}
{Bradley}, P.~A., \& {Winget}, D.~E., 1994, \apj, 430, 850

\bibitem[{Brickhill}(1992a)]{Brickhill92a}
{Brickhill}, A.~J., 1992a, \mnras, 259, 519

\bibitem[{Brickhill}(1992b)]{Brickhill92b}
{Brickhill}, A.~J., 1992b, \mnras, 259, 529

\bibitem[{Ising} \& {Koester}(2001)]{Ising01}
{Ising}, J., \& {Koester}, D., 2001, \aap, 374, 116

\bibitem[{Kupka}(1999)]{Kupka99}
{Kupka}, F., 1999, \apjl, 526, L45

\bibitem[{Kupka} \& {Montgomery}(2002)]{Kupka02}
{Kupka}, F., \& {Montgomery}, M.~H., 2002, \mnras, 330, L6

\bibitem[{Metcalfe}(2003)]{Metcalfe03}
{Metcalfe}, T.~S., 2003, \apjl, 587, L43

\bibitem[{Metcalfe} {et~al.}(2004)]{Metcalfe04}
{Metcalfe}, T.~S., {Montgomery}, M.~H., \& {Kanaan}, A., 2004, \apjl, 605, L133

\bibitem[{Metcalfe} {et~al.}(2000)]{Metcalfe00}
{Metcalfe}, T.~S., {Nather}, R.~E., \& {Winget}, D.~E., 2000, \apj, 545, 974

\bibitem[{Montgomery} \& {Kupka}(2004)]{Montgomery04a}
{Montgomery}, M.~H., \& {Kupka}, F., 2004, \mnras, 350, 267

\bibitem[{Montgomery} \& {Winget}(1999)]{Montgomery99a}
{Montgomery}, M.~H., \& {Winget}, D.~E., 1999, \apj, 526, 976

\bibitem[{Montgomery} {et~al.}(2003)]{Montgomery03}
{Montgomery}, M.~H., Metcalfe, T.~S., \& Winget, D.~E., 2003, \mnras, 344, 657

\bibitem[{Salpeter}(1961)]{Salpeter61}
{Salpeter}, E.~E., 1961, \apj, 134, 669

\bibitem[{Winget} {et~al.}(1991)]{Winget91} {Winget}, D.~E.,{Nather},
  R.~E., \& {Clemens}, J.~C., 1991, \apj, 378, 326
  
\bibitem[{Winget} {et~al.}(1994)]{Winget94} {Winget}, D.~E.,{Nather},
  R.~E., \& {Clemens}, J.~C., 1994, \apj, 430, 839

\bibitem[{Wu}(2001)]{Wu01}
{Wu}, Y., 2001, \mnras, 323, 248

\end{thebibliography}

\end{document}